\documentclass[%
 reprint,
 amsmath,amssymb,
 aps,
prl,
floatfix,
]{revtex4-1}

\usepackage[normalem]{ulem}
\usepackage{graphicx}
\usepackage{dcolumn}
\usepackage{bm}
\usepackage{amsfonts}
\usepackage{subeqnarray}
\usepackage{mathrsfs}
\usepackage{bbold}
\usepackage{appendix}
\usepackage{tikz}
\usetikzlibrary{arrows}
\usepackage{adjustbox}
\usepackage[caption=false]{subfig}
\begin{document}
\title{Theory of the supercurrent diode effect in  Rashba superconductors \\with arbitrary disorder}
\author{S. Ili\'{c} $^{1}$}
\author{F. S. Bergeret$^{1,2}$}

\affiliation{$^1$Centro de F\'isica de Materiales (CFM-MPC), Centro Mixto CSIC-UPV/EHU, Manuel de Lardizabal 5, E-20018 San Sebasti\'an, Spain}
\affiliation{$^2$Donostia International Physics Center (DIPC), Manuel de Lardizabal 4, E-20018 San Sebasti\'an, Spain}

%\date{\today}

\begin{abstract}
We calculate the non-reciprocal critical current and quantify the supercurrent diode effect in Rashba superconductors with arbitrary disorder, using the quasiclassical Eilenberger equation. The non-reciprocity is caused by the helical superconducting state, which appears when both inversion and time-reversal symmetries are broken.  In the absence of disorder, we find a very strong diode effect, with the non-reciprocity exceeding $40\%$ at  optimal temperatures, magnetic fields and spin-orbit coupling. We establish that the effect persists even in the presence of strong disorder. We show that the sign of the diode effect changes as  magnetic field and disorder are increased, reflecting the changes in the nature of the helical state.  
\end{abstract}
\maketitle
\paragraph{Introduction.-} The interplay between superconductivity, spin-orbit coupling (SOC),  and a Zeeman field leads to a variety of magnetoelectric effects widely studied in the past years \cite{szombati2016josephson,assouline2019spin,mayer2020gate,strambini2020josephson,edelstein1995,edelstein2005magnetoelectric,Buzdin:2008,mineev2011magnetoelectric,agterberg2012magnetoelectric,ojanen2012magnetoelectric,Konschelle:2015,Bergeret:2015}.   One of these effects is a nonreciprocal charge transport due to the breaking of time-reversal and inversion symmetries \cite{yasuda2019nonreciprocal,wakatsuki2017nonreciprocal,wakatsuki2018nonreciprocal,ando2020observation,baumgartner2021supercurrent, baumgartner2021effect, wu2021realization,he2021phenomenological,yuan2021supercurrent, daido2021intrinsic,bauriedl2021supercurrent,zinkl2021symmetry}.  Originally this effect was studied in the resistive regime, when $T\gtrsim T_c$, where superconducting fluctuations play a crucial role \cite{wakatsuki2017nonreciprocal,wakatsuki2018nonreciprocal}. More recently, it has been shown that nonreciprocity also manifests in the supercurrent in non-centrosymmetric superconducting structures  and in Josephson junctions \cite{dolcini2015topological, ando2020observation, baumgartner2021supercurrent, baumgartner2021effect, bauriedl2021supercurrent, he2021phenomenological,yuan2021supercurrent, daido2021intrinsic}. Specifically, the critical current depends on the direction of the current flow, and hence such systems are being suggested as superconducting diodes with  potential applications in low-power logic circuits. 

The non-reciprocity of the critical current can be quantified by the  superconducting  diode quality parameter
\begin{equation}
    \eta=(j_c^+-|j_c^-|)/(j_c^++|j_c^-|)\; , 
    \label{eq1}
\end{equation}
 where $j_c^{\pm}$ are the critical currents in  opposite directions. It has been shown in Refs. \cite{he2021phenomenological,yuan2021supercurrent, daido2021intrinsic} that $\eta$ is finite in noncentrosymetric superconducting systems in the presence of a  magnetic field. Namely, breaking of time-reversal and inversion symmetries in such systems allows for a formation of the helical superconducting phase, with the order parameter modulated in the direction transverse to the field: $\Delta(\mathbf{r})=\Delta e^{i\mathbf{q_0}\mathbf{r}}$. Therefore, Cooper pairs  have a finite momentum $\mathbf{q_0}$. Consequently, the depairing effect of supercurrents flowing parallel and anti-parallel to $\mathbf{q}_0$ is different, leading to a direction-dependent  critical current.

Importantly, the above-mentioned
 theoretical works   assume ideally pure superconducting structures and disregard the effect of disorder. However, the latter  is unavoidable in realistic structures, and therefore it is important to understand how  it  affects the supercurrent diode effect.  Moreover, understanding the role of disorder will enable to design devices based on combination of conventional materials.
 
In this Letter, we establish a microscopic theory of the supercurrent diode effect in disordered Rashba superconductors. As a base of our study, we use the quasiclassical equations for superconductors with strong Rashba SOC from Ref.~\cite{houzet2015quasiclassical}, which give a full description of the helical superconducting phase at arbitrary temperature, magnetic field and disorder. These equations provide a straightforward framework to discuss the diode effect, within which we readily reproduce previous results of numerical simulations in the ballistic limit \cite{ daido2021intrinsic}, and generalize them for arbitrary disorder. Moreover, we correct previous works based on the phenomenological Ginzburg-Landau (GL) theory ~\cite{he2021phenomenological, yuan2021supercurrent}, which overestimate the diode effect at weak fields and close to critical temperature.

 Our results elucidate the mechanisms leading to the diode effect, and show how it evolves in the full range of all relevant system parameters: SOC, magnetic field, temperature and disorder. Namely, the effect stems from the competition between two helical bands in a Rashba superconductor, which prefer opposite modulation vectors of the superconducting order parameter when magnetic field is applied. Both magnetic field and SOC are required for diode effect, however if either is too strong, the band competition ceases as one helical band begins to dominate, leading to the suppression of the effect. This means that a substantial $\eta$ exists only for some optimal magnetic field and SOC. Disorder further complicates this picture, as it introduces mixing of the two helical bands.  We discuss optimal parameter regimes where the effect is strongest (exceeding $\eta=40\%$ in the ballistic case). We establish that the effect persists at strong disorder, meaning that it can be realized even in disordered materials and hybrid systems. Moreover, we show that the sign $\eta$ changes as the magnetic field is increased \cite{daido2021intrinsic}, and also by increasing the disorder. The different signs can be related with different nature of the helical phase  at weak and strong field and disorder.

\paragraph{Quasiclassical theory.-}
The quasiclassical Eilenberger equation for a disordered Rashba superconductor with strong SOC, in the basis of two helical bands denoted with the index $\lambda=\pm1$, is \cite{houzet2015quasiclassical}:
\begin{equation}
v \mathbf{n} \cdot \boldsymbol{\partial}_\lambda \check{g}_\lambda+[(\omega+i \check{\Delta})\tau_z+\sigma_\lambda^{imp}, \check{g}_\lambda]=0.
\label{eq2}
\end{equation}
Here $\check{g}_\lambda$ is the quasiclassical Green's function in Matsubara representation, which is a  matrix in Nambu space spanned by the Pauli matrices $\tau_{x,y,z}$. $\omega=2\pi T(n+\frac{1}{2})$ is the Matsubara frequency, with $T$ being the temperature. Importantly, the two bands have the same Fermi velocity $v=\sqrt{2\mu/m+\alpha^2}$, but different densities of states, $\nu_\lambda=\nu(1-\lambda\alpha/v)$. Here, $\mu$ is the chemical potential, $m$ is the effective electron mass, $\alpha$ is the velocity associated with Rashba SOC, and $\nu=m/(2\pi)$. We introduced the derivative $\boldsymbol{\partial}_\lambda=\boldsymbol{\nabla}+i \frac{\lambda}{v} (\mathbf{h}\times \mathbf{z})[\tau_z,\cdot]$, $\mathbf{h}$ is the in-plane magnetic field, and $\mathbf{n}=\mathbf{p}/p_F=(n_x,n_y)$ describes the direction of the momentum at the Fermi level. The superconductivity is accounted by the term $\check{\Delta}=\Delta (\mathbf{r})\tau_++\Delta^*(\mathbf{r}) \tau_-$, with $\tau_{\pm}=\frac{1}{2}(\tau_x\pm i\tau_y)$, where $\Delta$ is the superconducting order parameter. The normalization condition is satisfied for each helical band: $\check{g}_\lambda^2=1$. 

Disorder is described by the self-energy $\sigma_\lambda^{imp}$ given as
\begin{equation}
\sigma_\lambda^{imp}=\sum_{\lambda'} (4 \tau_{\lambda'})^{-1}\big[\langle \check{g}_{\lambda'}\rangle+\lambda\lambda' \mathbf{n}\cdot \langle \mathbf{n}\check{g}_\lambda \rangle\big].
\label{eq3}
\end{equation}
Here, $\langle...\rangle$ denotes averaging over $\mathbf{n}$, and we introduced  $\tau_\lambda^{-1}=(1-\lambda \frac{\alpha}{v}) \tau^{-1}$, where $\tau^{-1}$ is the disorder scattering rate. Note that $\sigma_\lambda^{imp}$ stems from a simple scalar disorder potential, which acquires the form shown in Eq.~\eqref{eq3} upon projection to the helical basis~\cite{houzet2015quasiclassical}. 

Equation \eqref{eq2} is valid for $v\gtrsim \alpha$, as long as SOC is the dominant energy scale so that $\alpha p_F\gg \Delta, h, \tau^{-1}$. Under these conditions, interband pairing can be neglected, and Cooper pairs can be taken to be formed in each helical band separately. In the absence of disorder, the two bands are decoupled, while sharing the same superconducting gap $\Delta$. Any finite disorder mixes the two bands. 

To proceed, we assume that the superconducting phase varies only along the $x$-direction, and that the Zeeman field is applied along the $y$-direction: $\mathbf{h}=(0,h)$. We take that $\Delta(x)=\Delta e^{i q x}$
\footnote{\label{foot} Note that by taking the ansatz  $\Delta(x)=\Delta e^{i q x}$ we made an important assumption - that the helical phase appears in the whole phase diagram, at any $T$ and $h$.  This is, however, not always true. Namely, at $\alpha/v<0.25$, low $T$ and sufficiently high $h$, the so-called stripe phase, characterized by multiple modulation vectors,  can be stabilized instead of the helical phase \cite{dimitrova_theory_2007,agterberg2007magnetic}. Importantly, already at $\alpha/v=0.05$ the majority of the $h-T$ phase diagram is occupied by the helical phase with a small region of stripe phase. This region reduces by increasing $\alpha$, until it disappears at  $\alpha/v=0.25$ \cite{agterberg2007magnetic}. Disorder further suppresses the stripe phase. For these reasons, in this work we focus only on the helical phase, while neglecting the stripe phase. The effect this phase would have on the diode effect remains an interesting open question.} , where $q=q_0+\delta q$ is the phase gradient which contains two contributions: intrinsic modulation of the helical phase $q_0$, and an additional phase gradient caused by passing the supercurrent $\delta q$. Then, we write the Green's function as $\check{g}_\lambda(x)=g_\lambda \tau_z+\tilde{f}_\lambda(x) \tau_++\tilde{f}^*_\lambda (x) \tau_-$, and we may look for the solution in the form $\tilde{f}_\lambda(x)=- i f_\lambda e^{i q x }$. The normalization condition  gives $g_\lambda^2=1-f_\lambda^2$.  The Eilenberer equation then reduces to the following scalar equation
 \begin{multline}
f_\lambda (2\omega+in_x\rho_\lambda )=2 \Delta g_\lambda+
 \sum_{\lambda'}(2\tau_{\lambda'})^{-1} 
 \big[g_\lambda\langle f_{\lambda'}\rangle- f_\lambda\langle g_{\lambda'}\rangle  
 \\+\lambda \lambda' n_x \big(g_\lambda\langle n_x f_{\lambda'}\rangle -  f_\lambda\langle n_xg_{\lambda'}\rangle\big) \big].
\label{eq4}
\end{multline}
  Here, we introduced $\rho_\lambda=q v+2\lambda h$.

 The order parameter is determined self-consistently as
 \begin{equation}
\Delta\ln \frac{T}{T_c}+ \pi T\sum_{\omega>0}\sum_\lambda \bigg[\frac{\Delta}{\omega}-\bigg(1- \lambda \frac{\alpha}{v}\bigg) \langle f_\lambda \rangle \bigg]=0.
\label{eq5}
 \end{equation}
 Here, $T_c$ is the critical temperature of the superconductor in the absence of magnetic field. Finally,  the current along the $x$-direction is given as
 \begin{equation}
j=-4 \pi T i  v\sum_{\omega>0}\sum_{\lambda} \nu_\lambda \langle  n_x g_\lambda \rangle.
\label{eq6}
 \end{equation}
 Importantly, $j(q_0)=0$ - there should be no supercurrent flowing in the equilibrium~\cite{dimitrova_theory_2007}. 
 
Eqs.~\eqref{eq4},\eqref{eq5} and \eqref{eq6} are a starting point for studying the diode effect. First, $\Delta(q)$ should be calculated self-consistently from Eqs.~\eqref{eq4} and \eqref{eq5} for all values of $q$ where $\Delta$ is finite. Next, using $\Delta(q)$ obtained this way, one should calculate $j(q)$ from Eq.~\eqref{eq6}. Then, the critical currents in the two directions are determined as $j_c^+=\text{max}[j(q)]$ and $j_c^-=\text{min}[j(q)]$. Finally, the diode quality factor $\eta$ is obtained by replacing $j_c^\pm$ obtained this way in Eq.~\eqref{eq1}.

\paragraph{Ballistic case.-}
 
\begin{figure}[h!]
    \centering
    \includegraphics[width=0.35\textwidth]{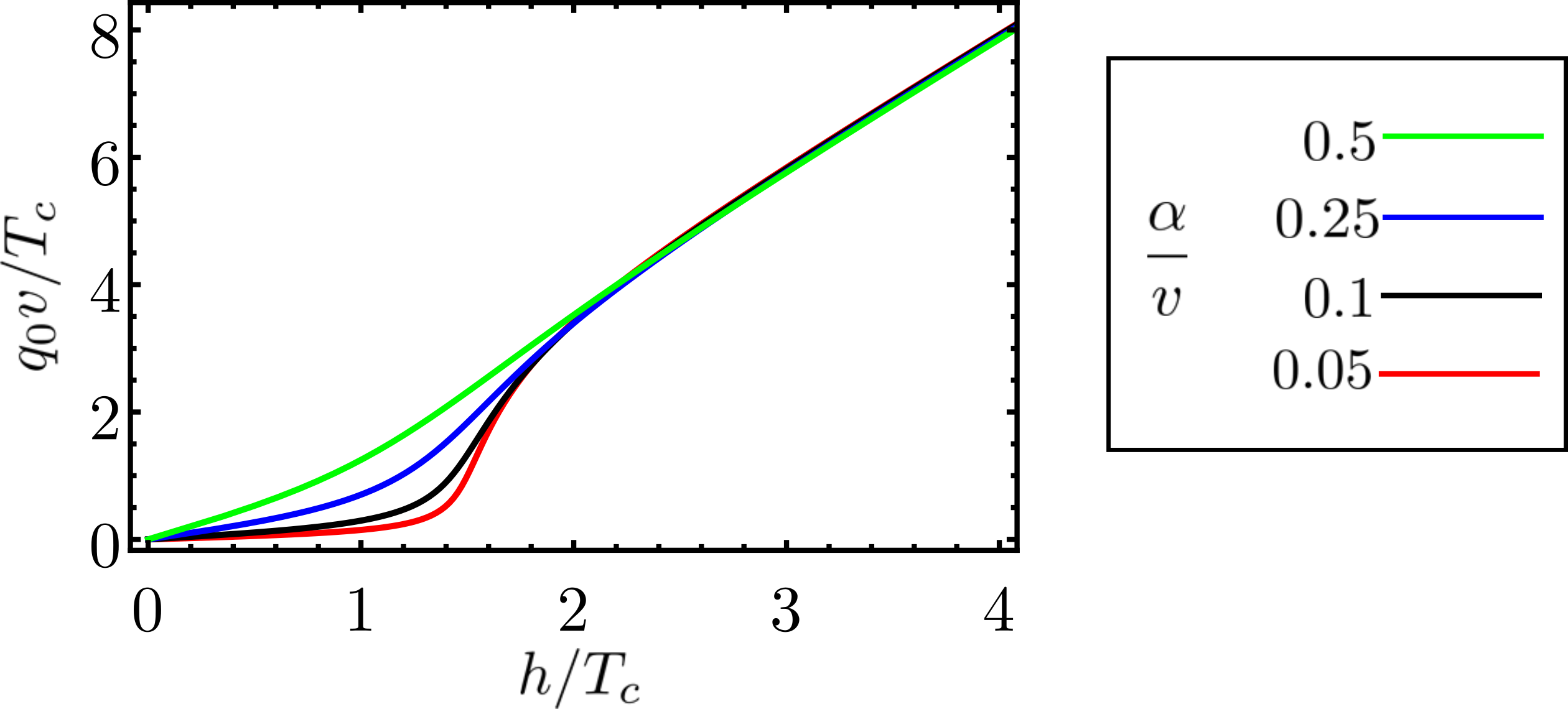}
    \caption{Helical modulation vector $q_0$ calcualted in the vicinity of the upper critical field $h_{c2}$, as a function of magnetic field, for different strengths of spin-orbit coupling. At low fields $q_0\approx 2\frac{\alpha}{v} h$ ("weak helical phase"), whereas at high fields $q_0\approx 2 h$ ("strong helical phase"). }
    \label{fig1}
\end{figure}

 Before discussing the diode effect, it is first useful to understand the evolution of the helical phase in magnetic fields. The two helical bands with $\lambda=\pm 1$ prefer opposite modulation vectors: $q_0^\lambda v=-2\lambda h$. At low magnetic fields, both bands contribute to helical superconductivity, which  yields a modulation vector $q_0 v\approx 2\frac{\alpha}{v}h$. This regime is known as a long-wavelength or "weak" helical phase \cite{dimitrova_theory_2007}. As the magnetic field is increased, the band with the higher density of states begins to dominate, whereas the contribution from the other band is suppressed. Therefore, at strong-enough field only one band contributes, and the modulation vector becomes $q_0v\approx2 h$. This is the short-wavelength of "strong" helical phase \cite{dimitrova_theory_2007}. The crossover from "weak" to "strong" phase is illustrated in Fig.~\ref{fig1}.
 Note that if the two bands have similar densities of states, at $\alpha/v\ll 1$, the so-called stripe phase might stabilize instead of the strong helical phase at high fields. The stripe phase is beyond the scope of the present work (see also the footnote [27]).

\begin{figure}[h!]
    \centering
    \includegraphics[width=0.4\textwidth]{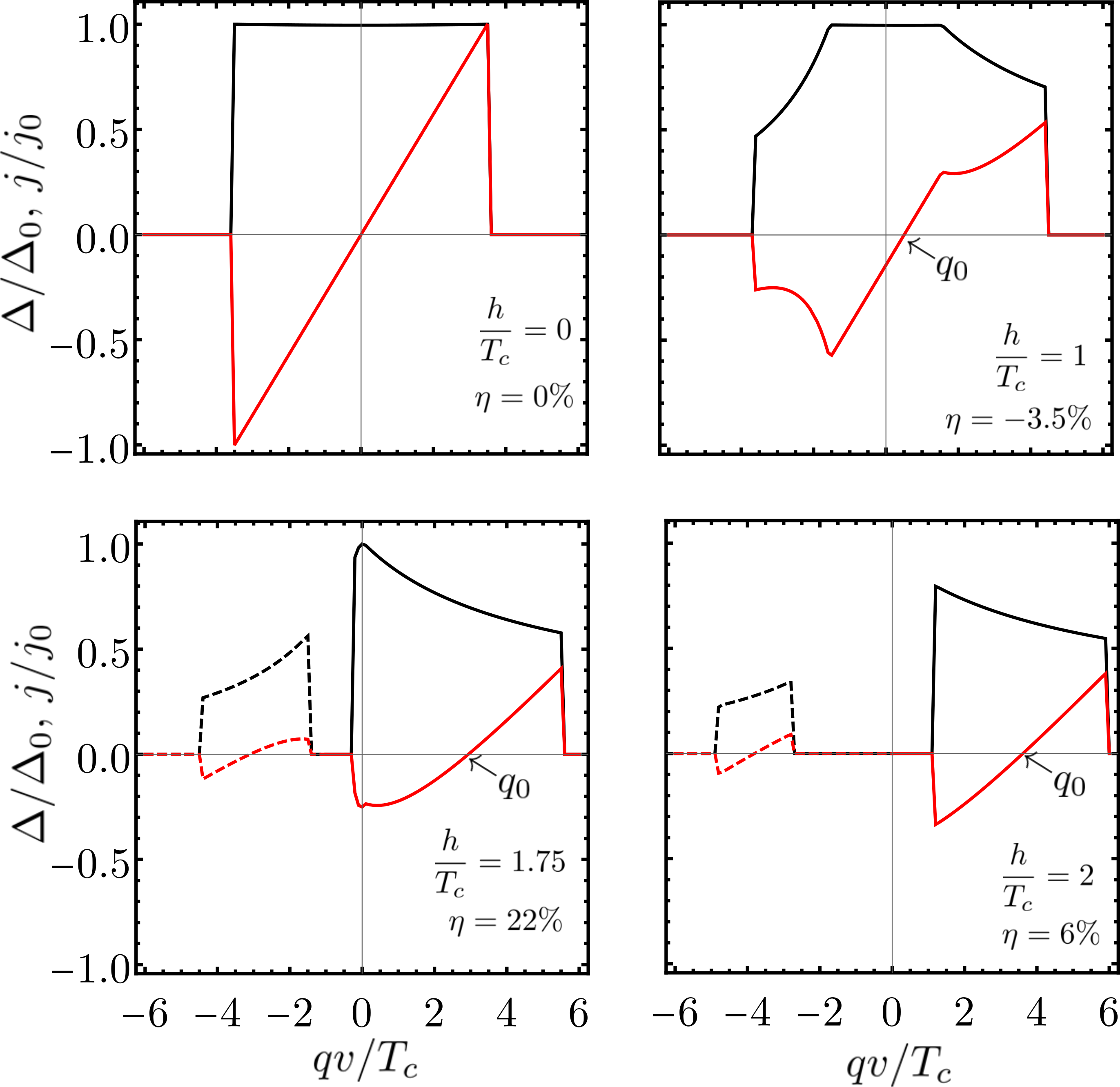}
    \caption{Superconducting gap (in black) and the supercurrent (in red), calculated self-consistently from Eqs.~\eqref{eq5} and \eqref{eq6}, respectively. Both quantities are plotted as a function of the phase gradient, for different values of the magnetic field. We normalize the curves with $\Delta_0$ and $j_0$, which are are the values of the superconducting gap and the critical current at $T=h=0$. We set $\alpha/v=0.25$ and $T=0.01T_c$. Note that at high fields (two lower panels), the self-consistency condition yields two solutions for $\Delta(q)$. The solution centered around $q_0$($-q_0)$, plotted with full (dashed ) line, predominantly comes from the helical band with higher (lower) density of states. The solution around $q_0$ is more stable (it minimizes the free energy \cite{agterberg2007magnetic}), and it is the only one relevant for our calculation.  }
    \label{fig2}
\end{figure}

 In the absence of disorder, the two helical bands are decoupled, and we readily find the solution of the Eilenberger equation \eqref{eq4} as $f_\lambda=2 \Delta/\mathcal{A}$ and $g_\lambda=(2 \omega+i n_x \rho_\lambda)/\mathcal{A}$, where $\mathcal{A}=\sqrt{(2\omega+i n_x \rho_\lambda)^2+4\Delta^2}$. The Fermi surface averages that enter  Eqs.~\eqref{eq5} and \eqref{eq6} can be found analytically (see the Supplementary Information). The critical current and the diode quality factor are then readily calculated following the procedure described below Eq.~\eqref{eq6}. Several examples of the self-consistent calculation of $\Delta(q)$, $j(q)$ and $\eta$ are shown in Fig.~\ref{fig2}. 

 The upper left panel of Fig.~\ref{fig2} shows the situation with $h=0$ and no helical phase. A phase gradient due to the supercurrent introduces depairing, and ultimately leads to a  phase transition to the normal state. The upper right panel of Fig.~\ref{fig2} depicts a situation where $h$ is sufficiently low so that the superconductor is in the "weak" helical state, whereas two lower panels depict a situation with the "strong" helical state. In these three panels, the current has a zero at  $q=q_0$ due to the intrinsic modulation of the helical phase, and all three show non-reciprocity of the critical current. The shape of $\Delta(q)$ and $j(q)$ in the "weak" and "strong" state is markedly different, leading to the different behavior of the diode effect. Namely, the effect is negative in the "weak" state ($j_c^+<|j_c^-|$, $\eta<0$) and positive in the "strong" state  ($j_c^+>|j_c^-|$, $\eta>0$).
 
\begin{figure}[h!]
    \centering
    \includegraphics[width=0.45\textwidth]{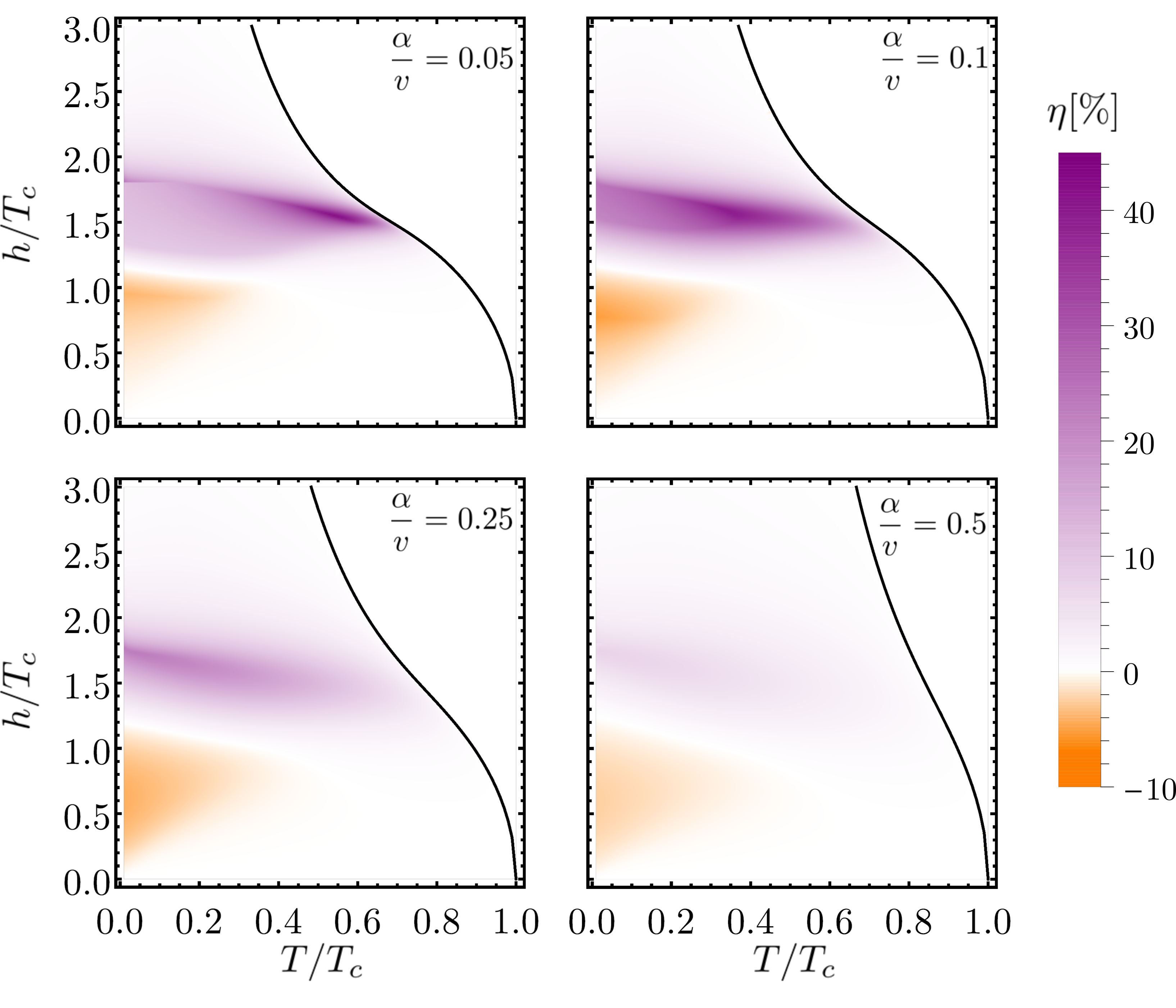}
    \caption{Diode quality factor $\eta$ for a ballistic superconductor, calculated for every point in the $h-T$ phase diagram at different strengths of spin-orbit coupling.}
    \label{fig3}
\end{figure}

 In Fig.~\ref{fig3}, we plot the diode quality factor $\eta$ for every point in the $h-T$ phase diagram for different strengths of SOC. The black curve in the plots correspons to the upper critical field $h_{c2}$.  At temperatures close to $T_c$  the diode effect is vanishingly small - we demonstrate this analytically up to linear order in $h$ using the GL theory in the Supplementary Information. This result is in contrast with Refs.~\cite{he2021phenomenological} and \cite{yuan2021supercurrent}, which don't take into account all relevant terms in the $q$-expansion of the GL free energy, and consequently find a finite effect in this regime.
 
 Fig.~\ref{fig3}  clearly illustrates the two regimes of the diode effect, driven by the "weak" and "strong" helical phases,  depicted with orange and purple colors, respectively. These results are in good qualitative agreement with the numerical study of Ref.~\cite{daido2021intrinsic}. Fig.~\ref{fig3} also showcases that the competition of two helical bands is a crucial ingredient for the diode effect. Namely, if one helical band becomes dominant, while the other one is fully suppressed, the diode effect disappears. One of the ways this can happen is by increasing the magnetic field - after the crossover to the "strong" phase, one band dominates. Another way is by increasing spin-orbit coupling - at large values of $\alpha/v$, one band will have much larger density of states then the other. In fact, at $\alpha/v=1$, only one helical band exists, and the diode effect disappears. Therefore, too large SOC and too large magnetic field both lead to the suppression of the diode effect, as illustrated in Fig.~\ref{fig3}.
  
\begin{figure}[h!]
    \centering
    \includegraphics[width=0.35\textwidth]{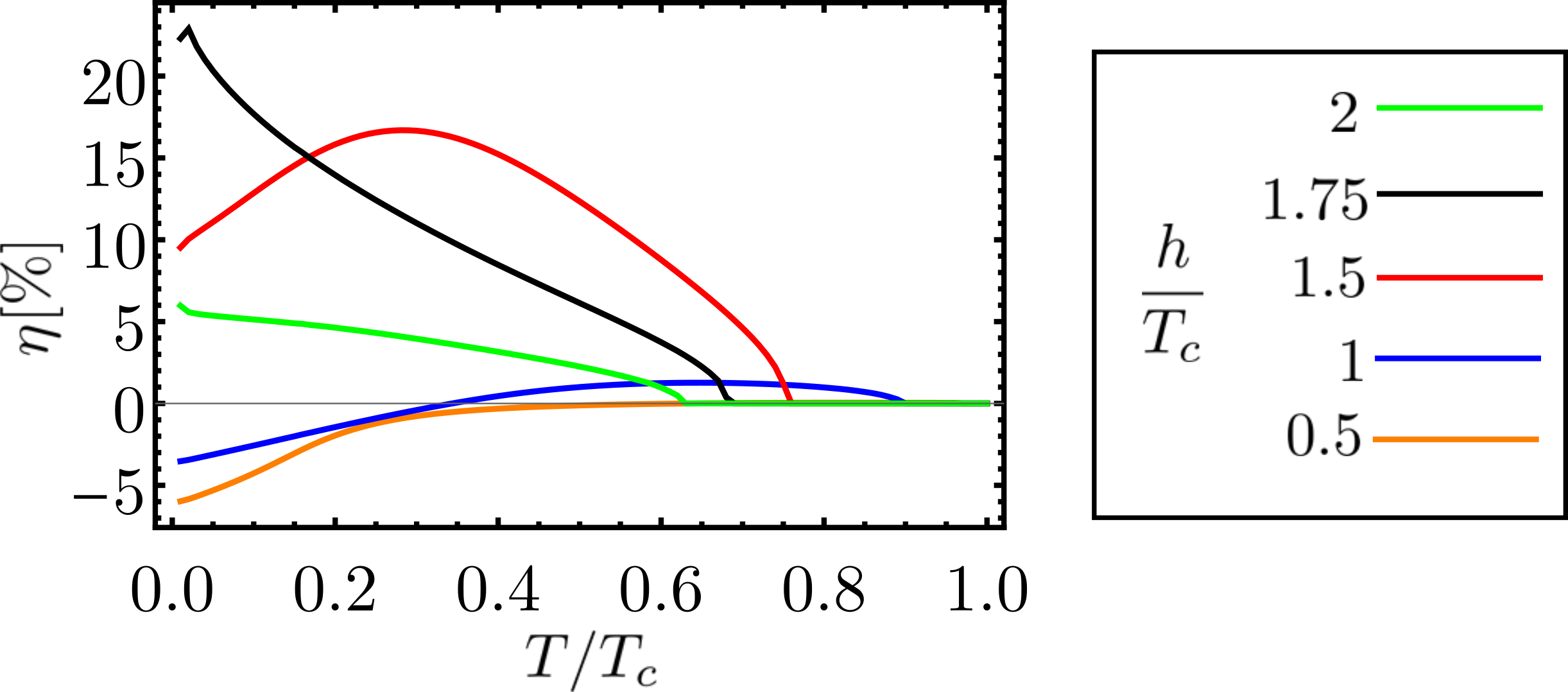}
    \caption{Diode quality factor as a function of temperature, for different values of the magnetic field, at $\alpha/v=0.25$.}
    \label{fig4}
\end{figure}

 Note that the coefficient $\eta$ can have non-monotonic dependence on the temperature at some fixed magnetic field, as shown in Fig.~\ref{fig4}. This can be explained by noticing that the diode effect is strongest close to the crossover field to the "strong" phase, combined with the fact that this field slightly reduces by increasing the temperature. Similar non-monotonic behavior of $\eta(T)$ was measured in a recent experiment in a few-layer NbSe$_2$\cite{bauriedl2021supercurrent}.

\begin{figure}[h!]
    \centering
    \includegraphics[width=0.35\textwidth]{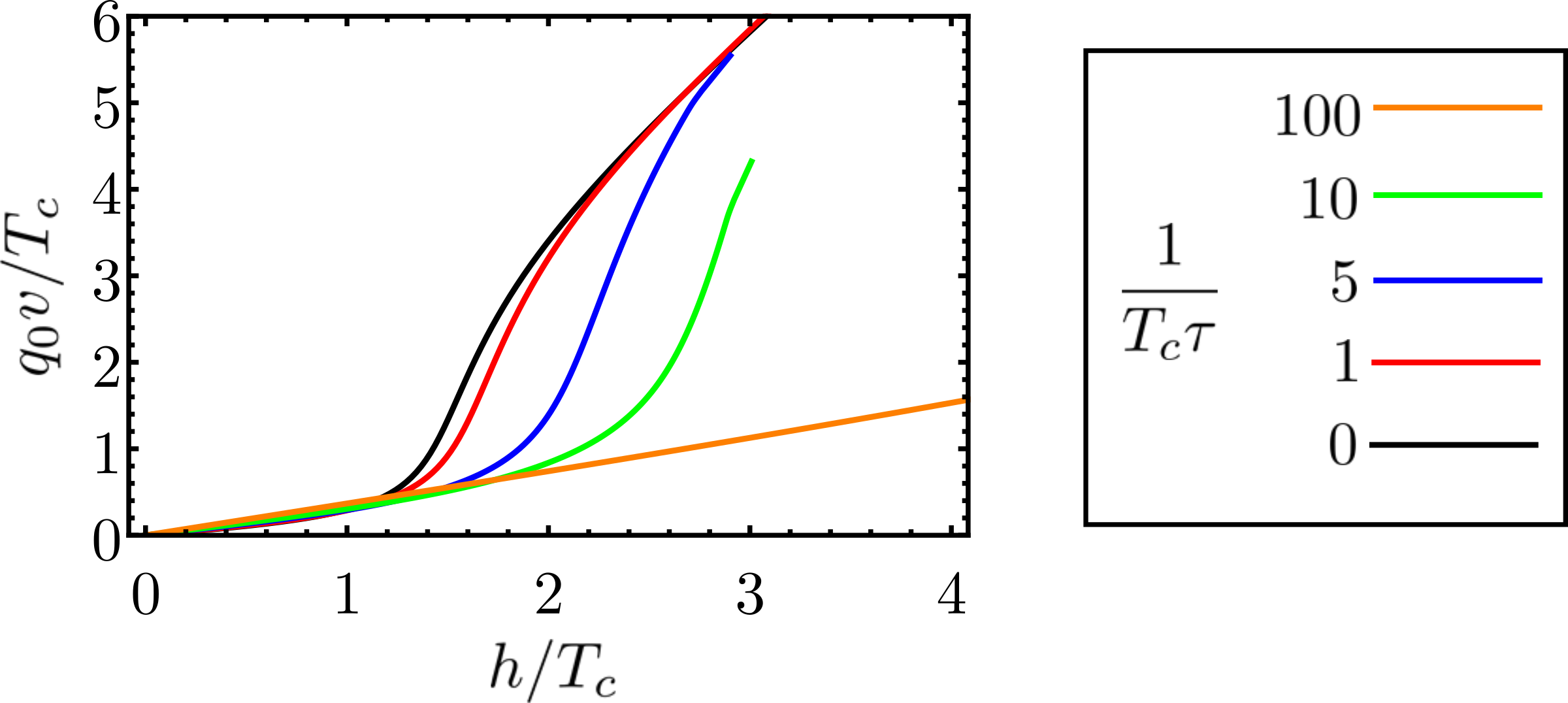}
    \caption{Helical modulation vector $q_0$ calculated in the vicinity of  $h_{c2}$, as a function of magnetic field, for different strenghts of disorder. We set $\alpha/v=0.1$. At strong disorder, $\tau^{-1}\gg \Delta,h $, we have $q_0\approx 4 \alpha h/(\alpha^2+v^2)$.}
    \label{fig5}
\end{figure}

\paragraph{Systems with disorder.-}
As established in the previous discussion in the ballistic limit, the competition between two helical bands upon applying the magnetic field is the driving force behind the diode effect. Very strong disorder mixes the bands, and therefore it suppresses this competition and the diode effect. 
By increasing disorder, the "strong" helical phase gets suppressed, and  for $\alpha p_F \gg \tau^{-1}\gg \Delta, h$, only the "weak" phase exists in the whole phase diagram, with the modulation vector  $q_0=4 \alpha h/(\alpha^2+v^2)$. This is illustrated in  Fig.~\ref{fig5}, where we plot $q_0$ for different values of disorder. 
In the following discussion, we explore the crossover from the strong diode effect in the ballistic case, to its vanishing at sufficiently strong disorder.

We examine the diode effect at arbitrary disorder in the GL regime, close to the phase transition to the normal state. This approach is valid for any $T$, as long as $h$ is sufficiently strong so that the $\Delta\ll T$.  In order to construct the GL free energy, we solve the Eilenberger equation \eqref{eq4} close to the phase transition. We may expand $f_\lambda$ up to third order in $\Delta$: $f_\lambda\approx f_\lambda^{(1)}+f_\lambda^{(3)}$, and $g_\lambda=\sqrt{1-f_\lambda^2}\approx 1-\frac{1}{2}\big(f_\lambda^{(1)}\big)^2$. The GL free energy is then
 \begin{equation}
F_q=\alpha_q  \Delta^2+\beta_q/2 \Delta^4
\label{eq8}
\end{equation}
where $ \alpha_q=\nu \ln \frac{T}{T_c}+2\pi T \sum_{\omega}\big[
\frac{\nu}{\omega}- \frac{1}{2\Delta}\sum_{\lambda}\nu_\lambda \big\langle f_\lambda^{(1)}\big\rangle
\big]$ and $\beta_q=-2\pi T \sum_\omega \sum_{\lambda} \frac{1}{2\Delta^3}\nu_\lambda \langle f_\lambda^{(3)}\big\rangle$ (see the Supplementary Information).
The order parameter is determined by minimizing the free energy with respect to $\Delta$, which gives
$\Delta^2=-\alpha_q/\beta_q$. From here,  we find the optimal free energy  $F_q^{opt}=-\alpha_q^2/(2\beta_q)$.
Finally, the current is given as
\begin{equation}
j=2 \partial_q F_q^{opt}.
\label{eq9}
\end{equation}

\begin{figure}[h!]
    \centering
    \includegraphics[width=0.45\textwidth]{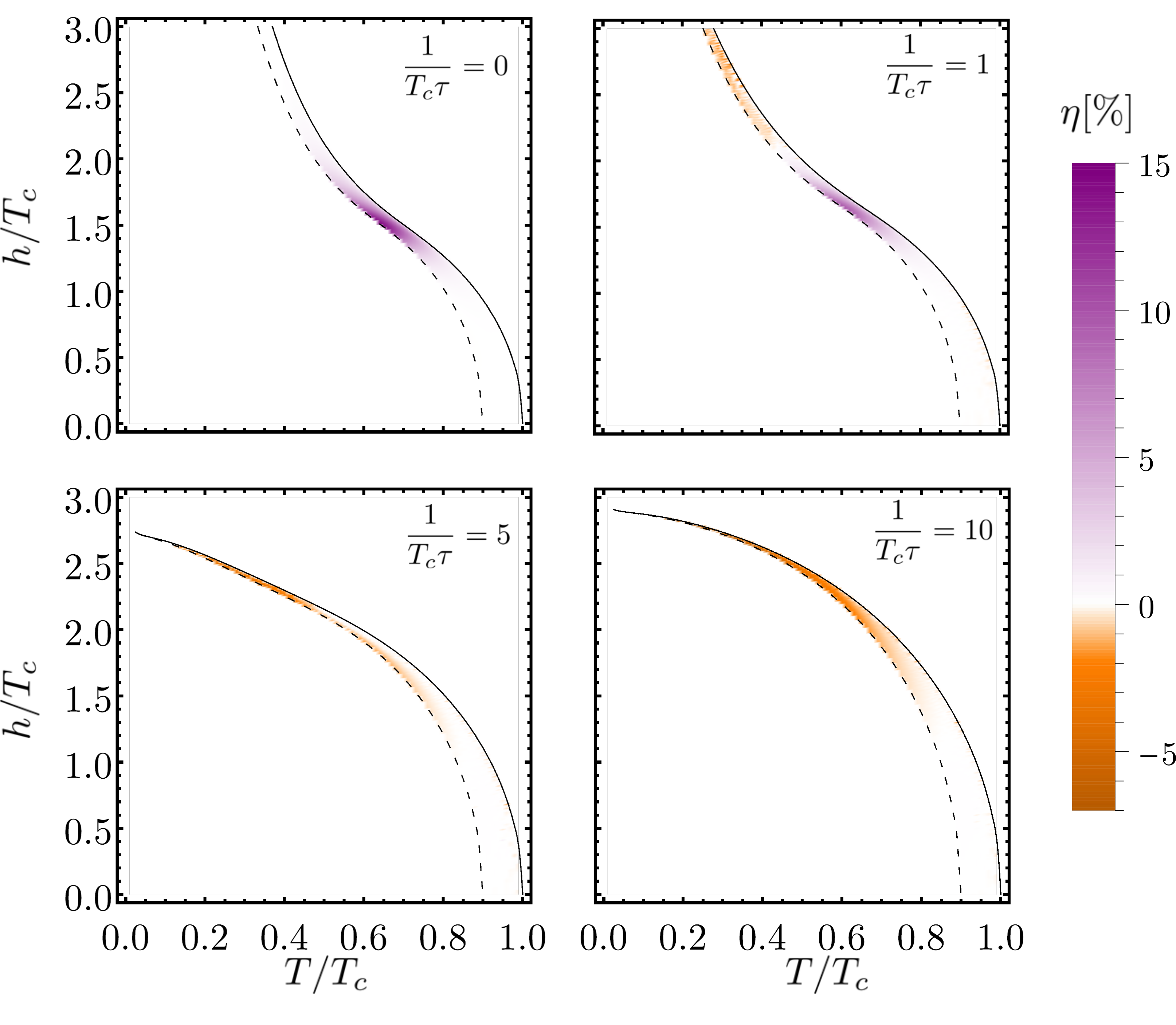}
    \caption{Diode quality factor in the disordered case, calculated in the GL regime at different strengths of disorder at $\alpha/v=0.1$. The full black line is the phase transition line $h_{c2}(T)$, whereas at the dashed line the condition $\Delta(T)=T$ is satisfied. Between the two lines $T>\Delta$  holds - this is the region of validity of the GL theory.   }
    \label{fig6}
\end{figure}

Fig.~\ref{fig6} shows the values of $\eta$ at different values of disorder calculated from Eq.~\eqref{eq9}. The upper left panel corresponds to the ballistic case, and agrees with the results of Fig.~\ref{fig3} obtained from the the full self-consistent calcualation. Notably, the diode effect qualitatively changes behavior as disorder is increased - it goes from positive to negative. This can be understood as a consequence of the crossover from the "strong" to the "weak" helical phase as disorder is increased, which correspond to $\eta>0$ and $\eta<0$, respectively, as established previously. The diode effect at $\tau^{-1}=10T_c$ reaches a sizeable value of $\eta\approx -7\%$.  Further increasing disorder ($\tau^{-1}>10 T_c$) leads to a qualitatively similar picture as in the lower right panel of Fig.~\ref{fig6}, but with smaller $\eta$. In the Supplementary Information,  we present an analysis of $\eta$ in a broader disorder range. For example, we find $\eta$ as large as $\sim -0.25\%$ at $\tau^{-1}=100 T_c$.  Note that the results obtained within the GL theory are only a lower bound of the effect, which likely reaches higher values beyond the GL regime.

\paragraph{Conclusion}
In summary, based on the quasiclassical formalism, we establish a theory of the diode effect in Rahsba superconductors with arbitrary disorder.  In the ballistic limit, we explore the whole phase diagram of helical superconductivity, and identify the conditions that maximize the diode effect. In the presence of disorder, we identify a new regime of the diode effect, which is qualitatively different from the ballistic limit, and show that a substantial diode effect exists even at strong disorder. Moreover, we show that the sign the quality factor $\eta$ is related to the nature of the helical state: $\eta>0$ in the "strong" state, and $\eta<0$ in the "weak" state. A possible device to experimentally study the  effect is a 2D superconductor with strong SOC in a heterostructure with a ferromagnetic insulator such as EuS \cite{hijano2021coexistence, strambini2017revealing,manna2020signature}, which induces a sizeable exchange field in the superconductor (a few Tesla) necessary to obtain a large diode effect. Moreover, such device is compatible with applications in superconducting electronics and spintronics, as it doesn't require applying external magnetic fields. 

\begin{acknowledgments}
      {\emph{Acknowledgements}}
  We thank Manuel Houzet for useful discussions. This work was supported by 
  European Union’s Horizon 2020 Research and Innovation Framework Programme under Grant No. 800923 (SUPERTED), and 
the Spanish Ministerio de Ciencia e Innovacion  (MICINN) through Project 
PID2020-114252GB-I00 (SPIRIT).
\end{acknowledgments}

\onecolumngrid
\pagebreak
\clearpage

\setcounter{equation}{0}
\setcounter{figure}{0}
\setcounter{table}{0}
\setcounter{page}{1}
\renewcommand{\thefigure}{S\arabic{figure}}
\renewcommand{\theequation}{S\arabic{equation}}
\begin{center}
	\textbf{\large Supplementary Information for "Theory of the supercurrent diode effect in Rashba superconductors with arbitrary disorder" }
\end{center}

This Supplementary Information contains technical details and additional analysis of the supercurrent diode quality factor $\eta$, which were omitted in the main text. 

In Sec.~\ref{Sec1}, we provide analytical expressions for the Fermi surface averages $\langle f_\lambda \rangle$ and $\langle n_x g_\lambda \rangle$ in the ballistic limit. The former is used in Eq.~(5) of the main text to self-consistently calculate the order parameter $\Delta$, whereas the latter is used to calculate the supercurrent $j$ in Eq.~(6) of the main text. In Sec.~\ref{Sec2}, we show how to solve the linearized Eilenberger equation, Eq.~(4) of the main text,  at arbitrary disorder. The solution is used to construct the Ginzburg-Landau (GL) free energy in Eq.~(7) of the main text. Using the free energy obtained this way, we demonstrate that the diode effect vanishes in the vicinity of $T_c$ up to first order in $h$ in Sec.~\ref{Sec3}. Finally, in Sec.~\ref{Sec4}, we examine the behavior of the diode quality factor $\eta$ for a broad range of disorder strength.

\section{Analytical evaluation of the Fermi surface averages in the ballistic limit \label{Sec1}}

In the ballistic limit, the Fermi surface averages $\langle f_\lambda \rangle$  and $\langle n_x g_\lambda \rangle$ can be expressed in terms of complete elliptic integrals of the first and second kind, namely  $K(x)=\int_0^{\pi/2} d\theta \sqrt{1-x \sin^2\theta}^{-1}$ and   $E(x)=\int_0^{\pi/2} d\theta \sqrt{1-x \sin^2\theta}$, respectively. We obtain
\begin{equation}
	\langle f_\lambda \rangle=\int_0^{2\pi} \frac{d\theta}{2\pi}\frac{2 \Delta}{\sqrt{(2\omega+i \cos \theta \rho_\lambda)^2+4\Delta^2}}= \frac{2\Delta}{\pi \sqrt{\Pi_\lambda^+}}K \bigg(\frac{8\Delta \rho_\lambda}{\Pi_\lambda^+}\bigg),
	\label{eqS1}
\end{equation}
\begin{multline}
	i\langle n_x g_\lambda \rangle=\int_0^{2\pi} \frac{d\theta}{2\pi} \frac{i (2\omega+i \cos\theta \rho_\lambda)\cos\theta}{\sqrt{(2\omega+i \cos \theta \rho_\lambda)^2+4\Delta^2}}=\frac{4\Delta}{\pi \sqrt{\Pi_\lambda^+}}K \bigg(\frac{8\Delta \rho_\lambda}{\Pi_\lambda^+}\bigg)
	-\frac{8 \Delta}{\pi \sqrt{\Pi_\lambda^-}}\text{Im} K \bigg(\frac{\Pi_\lambda^+}{\Pi_\lambda^-}\bigg) \\
	-\frac{\sqrt{\Pi_\lambda^-}}{\pi \rho_\lambda}\text{Im} E \bigg(\frac{\Pi_\lambda^+}{\Pi_\lambda^-}\bigg)
	+\frac{\sqrt{\Pi_\lambda^+}}{\pi \rho_\lambda}\text{Im} E \bigg(\frac{\Pi_\lambda^-}{\Pi_\lambda^+}\bigg)
	+\theta(\rho_\lambda) 
	\frac{2\sqrt{\Pi_\lambda^-}}{\pi \rho_\lambda }E\bigg(-\frac{8\Delta \rho_\lambda}{\Pi_\lambda^-}\bigg)
	-\theta(\rho_\lambda) 
	\frac{2\Pi_\lambda^+}{\pi \rho_\lambda \sqrt{\Pi_\lambda^-} }K\bigg(-\frac{8\Delta \rho_\lambda}{\Pi_\lambda^-}\bigg).
	\label{eqS2}
\end{multline}
Here, we used $n_x=\cos\theta$, and introduced $\Pi_\lambda^\pm=(2\Delta\pm \rho_\lambda)^2+4\omega^2$.
\section{Solution of the linearized Eilenberger equation at arbitrary disorder \label{Sec2}}
In order to construct the GL free energy introduced in Eq.~(7) of the main text, we solve the Eilenberger equation [Eq.~(4) ] close to the second-order phase transition to the normal state where $\Delta$ is small. First, we expand  $f_\lambda$ up to third order in $\Delta$: $f_\lambda\approx f_\lambda^{(1)}+f_\lambda^{(3)}$, and $g_\lambda=\sqrt{1-f_\lambda^2}\approx 1-\frac{1}{2}\big(f_\lambda^{(1)}\big)^2$. Here,  $f^{(1)}\sim \Delta$ and $f^{(3)}\sim \Delta^3$. Then, starting from Eq.~(4) of the main text, we find that the components $f^{(i)}$ ($i=1,3$)  satisfy the following equations
\begin{equation}
	f_\lambda^{(i)}\Omega_\lambda =X_\lambda^{(i)}+ \sum_{\lambda'} \bigg[
	\frac{\big\langle f_{\lambda'}^{(i)}\big \rangle}{2\tau_{\lambda'}}+\lambda \lambda' n_x \frac{\big \langle n_x f_{ \lambda'}^{(i)}\big \rangle}{2\tau_{\lambda'}}
	\bigg]
	.
	\label{eqS3}
\end{equation}
Here we introduced 
\begin{equation}
	\Omega_\lambda=2\omega+in_x \rho_\lambda+\tau^{-1}, 
	\label{eqS4}
\end{equation}
and the source terms 
\begin{equation}
	X_\lambda^{(1)}=2 \Delta, \qquad 
	X_\lambda^{(3)}=-\frac{\big(f_\lambda^{(1)}\big)^3}{2}\Omega_\lambda+\frac{f_\lambda^{(1)}}{2} \sum_{\lambda'} \bigg[
	\frac{\big\langle \big(f_{\lambda'}^{(i)}\big)^2\big \rangle}{2\tau_{\lambda'}}+\lambda \lambda' n_x \frac{\big \langle n_x \big(f_{ \lambda'}^{(i)}\big)^2\big \rangle}{2\tau_{\lambda'}}
	\bigg].
	\label{eqS5}
\end{equation}
To solve Eq.~\eqref{eqS3}, we take the averages:
\begin{equation}
	\big \langle f_\lambda ^{(i)}\big \rangle=\bigg\langle \frac{X_\lambda^{(i)}}{\Omega_\lambda} \bigg \rangle+
	\sum_{\lambda'} \bigg[\bigg \langle \frac{1}{\Omega_\lambda} \bigg \rangle 
	\frac{\big\langle f_{\lambda'}^{(i)}\big \rangle}{2\tau_{\lambda'}}+\lambda \lambda' \bigg \langle \frac{n_x}{\Omega_\lambda} \bigg \rangle \frac{\big \langle n_x f_{ \lambda'}^{(i)}\big \rangle}{2\tau_{\lambda'}}
	\bigg],
	\label{eqS6}
\end{equation}
\begin{equation}
	\big \langle n_x f_\lambda ^{(i)}\big \rangle=\bigg\langle \frac{n_x X_\lambda^{(i)}}{\Omega_\lambda} \bigg \rangle+
	\sum_{\lambda'} \bigg[\bigg \langle \frac{n_x}{\Omega_\lambda} \bigg \rangle 
	\frac{\big\langle f_{\lambda'}^{(i)}\big \rangle}{2\tau_{\lambda'}}+\lambda \lambda' \bigg \langle \frac{n_x^2}{\Omega_\lambda} \bigg \rangle \frac{\big \langle n_x f_{ \lambda'}^{(i)}\big \rangle}{2\tau_{\lambda'}}
	\bigg].
	\label{eqS7}
\end{equation}

Eqs.~\eqref{eqS6} and \eqref{eqS7} give two $4\times4$ coupled systems of equations, one for $(i)=(1)$ and the other for $(i)=(3)$, determining the averages $\big\langle f_\lambda^{(i)} \big \rangle$ and $\big\langle n_x f_\lambda^{(i)} \big \rangle$. First, we  solve Eqs.~\eqref{eqS6} and \eqref{eqS7} for $(i)=(1)$. Using $\big\langle f_\lambda^{(1)} \big \rangle$ and $\big\langle n_x f_\lambda^{(1)} \big \rangle$ obtained in this way, we next calculate $f_\lambda^{(1)}$ from Eq.~\eqref{eqS3}, which we then use to find the source term $X_\lambda^{(3)}$ from Eq.~\eqref{eqS5}. Using this source term, we finally solve Eqs.~\eqref{eqS6} and \eqref{eqS7} for $(i)=3$. This procedure allows us to find an analytical solution for the Green's functions at arbitrary disorder. However, it too  cumbersome so we do not write it here. 

In the absence of disorder, the solution significantly simplifies, and we have
\begin{equation}
	\langle f_\lambda^{(1)}\rangle= \frac{2\Delta}{\sqrt{\rho_\lambda^2+4\omega^2}}, \quad \langle f_\lambda^{(3)}\rangle=\frac{2\Delta^3(\rho_\lambda^2-8 \omega^2)}{(\rho_\lambda^2+4\omega^2)^{5/2}}.
\end{equation}

\section{Vanishing of the diode effect close to $T_c$ \label{Sec3}}
In this Section we demonstrate that the diode effect vanishes close to $T_c$ up to  linear order in  $h$, in contrast to previous works \cite{he2021phenomenological, yuan2021supercurrent}. For simplicity, in the following we focus only on the ballistic limit, but the same conclusion also holds at arbitrary disorder.  

\subsection{GL free energy close to $T_c$}
In the ballistic limit, the coefficients in the GL free energy, defined in Eq.~(7) of the main text,  are
\begin{equation}
	\alpha_q=\nu \ln \frac{T}{T_c}+2\pi T \nu \sum_{\omega>0}\bigg[
	\frac{1}{\omega}- \frac{1}{2}\sum_{\lambda=\pm 1}\bigg(1 -\lambda \frac{\alpha}{v}\bigg) \frac{2}{\sqrt{\rho_\lambda^2+4\omega^2}}
	\bigg], \quad 
	\beta_q=-2\pi T \nu \sum_\omega \sum_{\lambda=\pm 1} \frac{1}{2}\bigg(1 -\lambda \frac{\alpha}{v}\bigg)\frac{2(\rho_\lambda^2-8 \omega^2)}{(\rho_\lambda^2+4\omega^2)^{5/2}}.
\end{equation}
Close to $T_c$, we may expand $\alpha_q$ and $\beta_q$ assuming small $q$ and small $h$. We keep terms up to 4-th order in $q$ and up to first order in $h$. Then we have
\begin{equation}
	\alpha_q=-\alpha_0-\alpha_1 q+\alpha_2 q^2+\alpha_3 q^3-\alpha_4 q^4, \quad \beta_q=\beta_0+\beta_1 q-\beta_2 q^2-\beta_3 q^3+\beta_4 q^4.
	\label{eqS10}
\end{equation}
The coefficients in Eq.~\eqref{eqS10} are $\alpha_0=\nu\frac{T_c-T}{T_c}$, $\alpha_1=\frac{1}{2}\nu h\alpha C_3$, $\alpha_2=\frac{1}{8} \nu v^2 C_3$, $\alpha_3=\frac{3}{16} \nu h\alpha v^2 C_5$, $\alpha_4=\frac{3}{128} \nu v^4 C_5$, $\beta_0=\frac{1}{2} \nu C_3$, $\beta_1=\frac{3}{2} \nu h\alpha C_5$, $\beta_2=\frac{3}{8} \nu v^2 C_5$, $\beta_3=\frac{45}{32} \nu h\alpha v^2 C_7$, and $\beta_4=\frac{45}{256} \nu v^4 C_7$. 
Here, we introduced $C_m=2\pi T_c \sum_{\omega} 1/\omega^m,$ where $\omega=2\pi T_c (n+\frac{1}{2}).$

The optimal free energy is then
\begin{equation}
	F_q^{opt}=-\frac{1}{2}\frac{\alpha_q^2}{\beta_q}\approx -\frac{1}{2} \frac{\gamma_q^2}{\beta_0}.
	\label{eqS11}
\end{equation}
In order to simplify the following calculations, we introduced a new  quantity in Eq.~\eqref{eqS11} - $\gamma_q$, which combines the GL coefficients $\alpha_q$ and $\beta_q$, namely
\begin{equation}
	\gamma_q=\frac{\alpha_q}{\sqrt{1+\frac{\beta_1 q}{\beta_0}-\frac{\beta_2q^2}{\beta_0}-\frac{\beta_3 q^3}{\beta_0}+\frac{\beta_4 q^4}{\beta_0}}}\approx 
	(-\alpha_0-\alpha_1 q+\alpha_2 q^2+\alpha_3 q^3-\alpha_4 q^4)\bigg(1-\frac{1}{2} \frac{\beta_1}{\beta_0} q+\frac{1}{2} \frac{\beta_2}{\beta_0} q^2+\frac{1}{2} \frac{\beta_3}{\beta_0} q^3-\frac{1}{2} \frac{\beta_4}{\beta_0}q^4 \bigg).
	\label{eqS12}
\end{equation}
To proceed, we keep terms up to $q^4$ in $\gamma_q$ and in the leading order in $\alpha_0$. This way, we obtain 
\begin{equation}
	\gamma_q=-\gamma_0-\gamma_1 q+\gamma_2 q^2-\gamma_3 q^3+\gamma_4 q^4.
	\label{eqS13}
\end{equation}
The coefficients $\gamma_i$ are
\begin{align}
	&\gamma_0=\alpha_0=\nu \frac{T_c-T}{T_c}, \qquad \gamma_1=\alpha_1=\frac{1}{2}\nu h\alpha C_3, \quad \gamma_2=\alpha_2=\frac{1}{8} \nu v^2 C_3,  \nonumber \\
	&\gamma_3=-\alpha_3+\frac{1}{2}\frac{\alpha_1\beta_2}{\beta_0}+\frac{1}{2}\frac{\alpha_2\beta_1}{\beta_0}=\frac{3}{16} \nu C_5 h v^2 \alpha, \nonumber   \\
	& \gamma_4=-\alpha_4+\frac{1}{2}\frac{\alpha_2\beta_2}{\beta_0}=\frac{3}{128} \nu C_5 v^4.
	\label{eqS14}
\end{align}
As we show in the next section, it is crucial to keep terms up to $q^4$ to describe correctly the diode effect.  Such terms have been neglected in Ref.~\cite{yuan2021supercurrent}, while in Ref.~\cite{he2021phenomenological} they have been considered but treated inconsistently,  which  lead to an incorrect prediction of a finite diode effect linear in $h$ within the GL regime.
\subsection{Supercurrent and the diode effect close to $T_c$}
The supercurrent can be calculated from the free energy as
\begin{equation}
	j=2\partial_q F_q^{opt}=-\frac{2}{\beta_0}\gamma_q \partial_q \gamma_q,
	\label{eqS15}
\end{equation}
which yields
\begin{equation}
	j=-\frac{2}{\beta_0} (-\gamma_0-\gamma_1 q+\gamma_2 q^2-\gamma_3 q^3+\gamma_4 q^4)(-\gamma_1 +2 \gamma_2 q-3\gamma_3 q^2+4\gamma_3 q^3).
	\label{eqS16}
\end{equation}
The critical current $j_c$ is found at the critical momenta $q_c$, determined from $\frac{d j}{d q}\bigg|_{q=q_c}=0$. 
Solving this equation perturbatively, up to first order in $h$ and $\gamma_4$, we obtain
\begin{equation}
	q_c^{\pm}=\pm \frac{1}{\sqrt{3}}\sqrt{\frac{\gamma_0}{\gamma_2}}+\frac{\gamma_1}{2\gamma_2}+\frac{ \gamma_3\gamma_0}{18 \gamma_2^2}-\frac{\gamma_0\gamma_1\gamma_4}{9 \alpha_2^3}.
	\label{eqS17}
\end{equation}
Replacing Eq.~\eqref{eqS17} into Eq.~\eqref{eqS16}, we obtain the critical current
\begin{equation}
	j_c^{\pm}=j(q_c^\pm)=\frac{8}{9 \gamma_2^2\beta_0}\bigg[
	\pm \sqrt{3}\gamma_0^{3/2}\gamma_2^{5/2}+\gamma_0^2(2\gamma_1\gamma_4-\gamma_2\gamma_3)
	\bigg].
	\label{eqS18}
\end{equation}
The diode quality factor is then
\begin{equation}
	\eta=\frac{j_c^+-|j_c^-|}{j_c^++|j_c^-|}=\frac{\sqrt{\gamma_0}}{\sqrt{3}\gamma_2^{5/2}}(2\gamma_1\gamma_4-\gamma_2\gamma_3).
	\label{eqS19}
\end{equation}
Finally, we use the coefficients given in Eq.~\eqref{eqS14}
\begin{equation}
	\eta\propto 2\gamma_1\gamma_4-\gamma_2\gamma_3=h\alpha C_3 C_5 v^4 \bigg(\frac{3}{128}-\frac{3}{128}\bigg)=0.
	\label{eqS0}
\end{equation}
The diode quality factor therefore vanishes close to $T_c$ up to first order in $h$. 

\section{Dependence of $\eta$ on disorder strength \label{Sec4}}
\begin{figure}[h!]
	\centering
	\includegraphics[width=0.99 \textwidth]{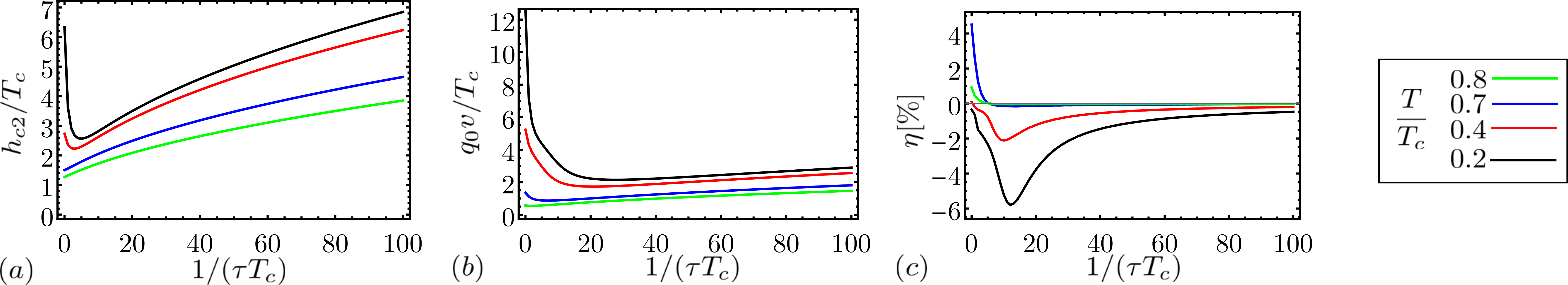}
	\caption{ $(a)$ and $(b)$: Upper critical field $h_{c2}$ [panel $(a)$] and the corresponding helical modulation vector $q_0$ [panel $(b)$], plotted as a function of disorder at various temperatures. $(c)$: Diode quality factor $\eta$ as a function of disorder, for various temperatures. $\eta$ is calculated along the $h_{c2}$ curve presented in panel $(a)$, at a field $h$ slightly below $h_{c2}$ so that $\Delta(h,q_0)=0.2 T_c$. }
	\label{figsup}
\end{figure}

In this Section, we examine the behavior of the diode quality factor $\eta$ in a broad range of disorder strength, as shown in Fig.~\ref{figsup}. To do so we use the GL formalism [Eqs.~(7) and (8) of the main text].  First, in panels $(a)$ and $(b)$, we calculate the upper critical field $h_{c2}$ and the corresponding helical modulation vector $q_0$ at a fixed temperature as a function of disorder. These quantities are calculated by imposing the condition of the second-order phase transition $\alpha_q=0$, together with the condition of zero current $\partial_q\alpha_q=0$. Then, in panel $(c)$, we calculate $\eta$ as a function of disorder along the $h_{c2}({\tau})$ curve shown in panel $(a)$.

The results of panel $(c)$ clearly illustrates the two different regimes of the diode effect established in the main text: positive $\eta$ at weak disorder (most pronounced in the blue curve),  and a negative $\eta$ at strong disorder (most pronounced in the red and black curves). 
Moreover, we see that $\eta$ has a substantial value in a broad range of disorder, meaning that this effect can be expected even in very disordered materials and structures. Note that the results presented here provide only a lower bound for $\eta$ in the GL regime - $\eta$ is likely larger away from the $h_{c2}(T)$ transition line. To determine  $\eta$ for arbitrary disorder and low temperatures one has to  solve numerically Eqs.~(4) of the main text, and to use the solution in Eqs.~(5) and (6) of the main text in order to find $\Delta$ and $j$ self-consistently. This is beyond the scope of the present work.

\end{document}